\documentclass[aps,prb,twocolumn,superscriptaddress,showpacs,floatfix]{revtex4}

\newcommand{\Ef}{E_{f}}
\newcommand{\Ep}{E^{\prime}}
\newcommand{\Vo}{V_{0}}
\newcommand{\Eq}{E(q)}
\newcommand{\omo}{\omega}
\newcommand{\ko}{k}
\newcommand{\kv}{\kappa}
\newcommand{\kvp}{\kappa_{\mathrm{p}}}

\newcommand{\vrr}{\mathbf{r}}
\newcommand{\vru}{\hat{\mathbf{r}}}
\newcommand{\vxu}{\hat{\mathbf{x}}}
\newcommand{\vx}{\mathbf{x}}

\newcommand{\vy}{\mathbf{y}}
\newcommand{\vq}{\mathbf{q}}

\newcommand{\vz}{\mathbf{z}}
\newcommand{\vp}{\mathbf{p}}
\newcommand{\vpu}{\hat{\mathbf{p}}}
\newcommand{\puxt}{\hat{p}_{x}^{2}}
\newcommand{\puyt}{\hat{p}_{y}^{2}}
\newcommand{\puzt}{\hat{p}_{z}^{2}}
\newcommand{\vE}{\mathbf{E}}
\newcommand{\vEvac}{\mathbf{E}^{\mathrm{v}}}
\newcommand{\Gvac}{\Gamma^{\mathrm{v}}}
\newcommand{\Svac}{\Sigma^{\mathrm{v}}}
\newcommand{\Go}{\Gamma^{0}}
\newcommand{\ga}{\gamma}
\newcommand{\Bn}{B_{n}}
\newcommand{\Bnd}{B_{n}^{\dag}}
\newcommand{\Bm}{B_{m}}
\newcommand{\Bq}{B_{q}}
\newcommand{\Bqd}{B_{q}^{\dag}}

\newcommand{\Rep}{\mathrm{Re}}
\newcommand{\Imp}{\mathrm{Im}}

\newcommand{\ct}{\cos\theta}
\newcommand{\cst}{\cos^{2}\theta}
\newcommand{\sst}{\sin^{2}\theta}
\newcommand{\Vnvac}{\Sigma^{\mathrm{v}}}
\newcommand{\Vns}{\Sigma^{\mathrm{s}}}
\newcommand{\Vn}{\Sigma}
\newcommand{\Vxs}{\Sigma^{\mathrm{s}}(k,x)}
\newcommand{\Wxs}{\Sigma^{\mathrm{s}}(k,x)}
\newcommand{\Vxsx}{\Sigma^{\mathrm{s}}_{\mathbf{x}}(k,x)}
\newcommand{\Vxsy}{\Sigma^{\mathrm{s}}_{\mathbf{y}}(k,x)}
\newcommand{\Vxsz}{\Sigma^{\mathrm{s}}_{\mathbf{z}}(k,x)}
\newcommand{\Vqsz}{\Sigma^{\mathrm{s}}_{\mathbf{z}}(k,q)}

\newcommand{\Dkq}{\Sigma (k,q)}
\newcommand{\Dqs}{\Sigma^{\mathrm{s}}(k,q)}
\newcommand{\Dqsx}{\Sigma^{\mathrm{s}}_{\,\mathbf{x}}(q)}
\newcommand{\Dqsy}{\Sigma^{\mathrm{s}}_{\,\mathbf{y}}(q)}
\newcommand{\Dqsz}{\Sigma^{\mathrm{s}}_{\,\mathbf{z}}(q)}
\newcommand{\Qo}{Q_{0}}
\newcommand{\Ko}{K_{0}}
\newcommand{\Kone}{K_{1}}
\newcommand{\eps}{\epsilon}
\newcommand{\epsp}{\eps^{\prime}}
\newcommand{\epspp}{\eps^{\prime\prime}}
\newcommand{\kvpp}{\kvp^{\prime}}
\newcommand{\kvppp}{\kvp^{\prime\prime}}
\newcommand{\gae}{\gamma_{\eps}}
\newcommand{\vEs}{\mathbf{E}^{\mathrm{s}}}

\usepackage{graphicx}
\usepackage{amsmath}
\usepackage{latexsym}
\begin{document}

\title{Excitons in long molecular chains near the reflecting interface}

\author{Yu.~N.~Gartstein}
\affiliation{Department of Physics, The University of Texas at
Dallas, Richardson, TX 75083, USA}
\author{V.~M.~Agranovich}
\affiliation{UTD-NanoTech Institute, The University of Texas at
Dallas, Richardson, TX 75083, USA} \affiliation{Institute of
Spectroscopy, Russian Academy of Science, Troitsk, Moscow, Russia}


\begin{abstract}
We discuss coherent exciton-polariton states in long molecular
chains that are formed due to the interaction of molecular
excitations with both vacuum photons and surface excitations of
the neighboring reflecting substrate. The resonance coupling with
surface plasmons (or surface polaritons) of the substrate can
substantially contribute to the retarded intermolecular
interactions leading to an efficient channel of the decay of
one-dimensional excitons with small momenta via emission of
surface excitations. The interface also modifies the radiative
decay of excitons into vacuum photons. In an idealized system,
excitons with higher momenta would not emit photons nor surface
waves. For a dissipative substrate, additional exciton quenching
takes place owing to Joule losses as the electric field of the
exciton polarization penetrates the substrate. We discuss how
these effects depend on the polarization of molecular excitations,
their frequency and on the distance of the chain from the
substrate.
\end{abstract}


\maketitle

\section{Introduction.}

The interaction of the electromagnetic field with a molecular
excitation in an aggregate of identical molecules leads, on the
one hand, to the delocalization of the excitation over the
aggregate and, on the other hand, to a modification of its
radiative decay.\cite{VMAbook} Both notions of excitons and
polaritons are used in the literature on such delocalized
excitations. In crystalline structures, the excitation can be
spatially coherent and then it is characterized by its wave vector
as a proper quantum number. In this paper we discuss how
electric-dipole-active coherent excitations in linear crystals are
affected by the presence of a neighboring metallic/dielectric
half-space.

A variety of one-dimensional (1D) electronic systems available
nowadays, such as conjugated polymers, $J-$ and $H-$ aggregates,
semiconducting quantum wires and carbon nanotubes, exhibit
interesting optical properties and are considered for potential
applications in optoelectronics; their spectroscopy is an active
research area. Successes at the synthesis and fabrication of these
systems have resulted in the continuously improving quality and
the increase of their ``conjugation length'' $L$ which may exceed
the appropriate electromagnetic wave length $\lambda$. Perhaps one
of the most noteworthy achievements in this regard is a recent
observation \cite{PDAnature} of a macroscopic coherence of a
single exciton in polydiacetylene chains of $L \simeq 10$ $\mu$m
that allowed to discuss an issue of ``an ideal 1D quantum wire''.
\cite{PDAbaessler} It is also relevant to note a physically very
different but conceptually related class of excitations in
 chains of ``dipole-coupled'' nanoparticles (see, e.~g., a recent
 Ref.~\onlinecite{marksar}
and multiple references therein) studied for photonic and
plasmonic applications.

Long before the modern experimental advances, it was shown
\cite{AD,VMAbook} that the coherent interaction of low-dimensional
(1D and 2D) excitons with the electromagnetic field is drastically
different than in 3D systems. The radiative decay of
low-dimensional excitons is strongly enhanced in the region of
their wave vectors $|\vq| \leq k$, where $k=\omo/c$ ($c$ is the
speed of light in vacuum) is determined self-consistently by the
excitation energy $E(\vq)=\hbar \omo$, while the excitations with
$|\vq|
> k$ would not radiate, as is required by the energy and momentum
conservation for an exciton-photon system. More specifically, for
1D excitons in the molecular chain in vacuum, the radiative width
$\Gamma=\hbar/\tau$ ($\tau$ being the decay time) depends on the
wave vector $q$ as
\begin{eqnarray}
\Gvac(k,q) & = & \frac{\pi p^2}{a} \left[ 2(k^2-q^2)\cst \nonumber
\right. \\
& + & \left.  (k^2+q^2)\sst
 \right] \Theta(k-|q|), \label{radvac}
\end{eqnarray}
where $p$ is the magnitude of the molecular dipole transition
moment $\vp$ that makes angle $\theta$ with the chain axis, $a$
the intermolecular spacing along the chain, and $\Theta$ the
step-function. As compared with the radiative width
\begin{equation}\label{radmol}
\Go(k)=4p^2 k^3/3
\end{equation}
of a single molecule, Eq.~(\ref{radvac}) exhibits an enhancement
factor of $\sim \lambda/a=2\pi/k a \sim 10^2-$10$^3$ in the
optical region of the spectrum (``superradiant'' states). This
enhancement has been discussed in the context of various systems
(see, e.~g.,
Refs.~\onlinecite{phil1975b,citrin1992,chen2001,spataru2005} and
references therein). Equation (\ref{radvac}) also shows how the
polarization of the transition dipole affects the $q-$dependence
of the decay rate.

\begin{figure*}
\includegraphics[scale=0.75]{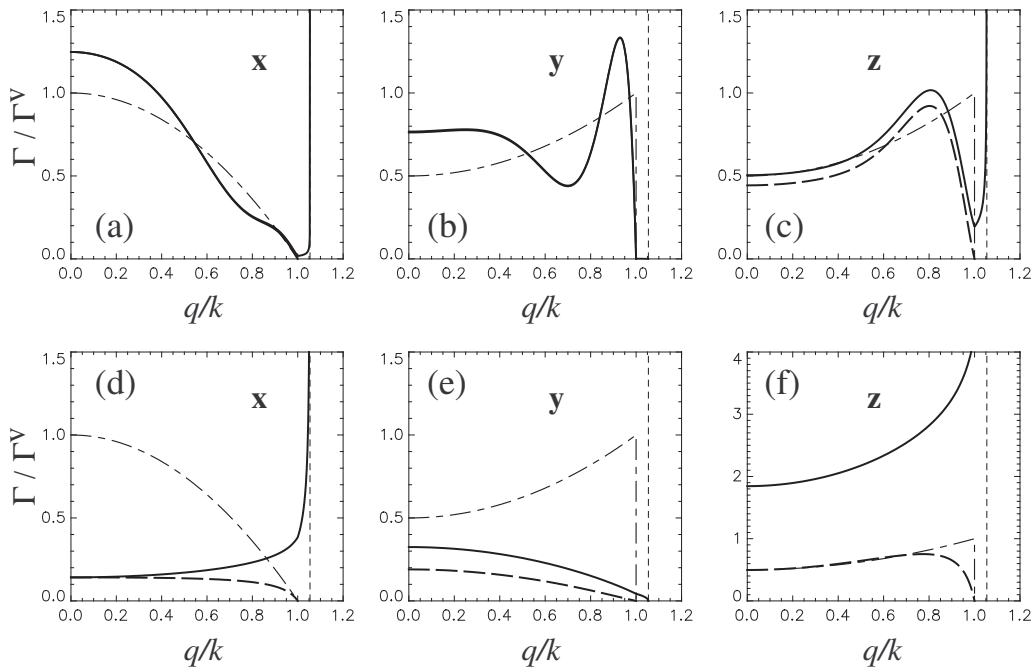}
\caption{\label{im_10_0}Decay width $\Gamma (k,q)$ of 1D
exciton-polaritons as a function of the reduced wave vector $q/k$
for a fixed value of $k=2\pi/400\, a$ and substrate's dielectric
constant $\eps=-10$. Two rows correspond to different
dipole-to-image-dipole distances $d$: upper panels (a--c) are for
$d=600\, a$, lower panels (d--f) for $d=10\, a$. All distances are
measured in units of intermolecular spacing in the chain, $a$.
Each of the trio of the panels in a row corresponds to different
exciton polarization, indicated by the boldface letters. The chain
is parallel to the $x$-axis and situated above the substrate whose
surface is the $xy$-plane. Decay width is shown with respect to
$\Gamma^{\mathrm{v}}=2\pi p^2 k^2/a$ which is the radiative width
of the $\vx$-polarized exciton in vacuum at $q=0$,
Eq.~(\ref{radvac}). The overall vacuum benchmark results
(\ref{radvac}) are shown with the dash-dotted lines. The total
decay width is displayed by thick solid lines, the dashed solid
lines (when distinguishable) show the part of the width due to the
decay into vacuum photons only. The vertical dash lines indicate
the position of the surface plasmon wave vector, $\kvp/k$,
Eq.~(\ref{SPP}).}
\end{figure*}

Importantly for applications, it is possible to manipulate the
optical properties of molecular excitations and to form new hybrid
excitations by putting molecules or molecular aggregates in the
vicinity of interfaces and in dielectric
microcavities.\cite{Novbook} Well-known, e.g., is an oscillating
dependence of the radiative width on the distance of a single
molecule from the planar interface resulting from the interference
of the radiative fields of a molecular transition and its image.
\cite{cps1978,Novbook} The dipole-dipole interaction gets also
modified in the vicinity of a surface or in the cavity.
\cite{silbey1995,stuart1998,hartman2001} In the case of a linear
molecular crystal, the environment can lead to qualitatively
interesting coherent effects as it is the interaction of many
molecular transition dipoles (and their images) that would
determine the properties of the excitation. Philpott,
\cite{phil1975b} for instance, pointed out that by placing a
linear chain near a transparent substrate, one could probe some of
otherwise non-radiant polariton states via emission of bulk
substrate photons. It was also studied how the radiative decay
properties of quantum wire excitons get modified when the wire is
embedded in a microcavity, that is, via emission of cavity photons
(Refs.~\onlinecite{chen2001,chen2001a} and references quoted
there). As one-dimensional arrays are considered for the directed
energy transfer applications, their interaction with the
environment may also be used to achieve certain purposes such as,
e.g., to counteract losses by embedding the array in the gain
medium.\cite{citringain}

In this paper we discuss 1D coherent dipole excitations that are
formed in the neighborhood of the planar reflecting substrate in
the range of frequencies $\omo$ where the dielectric constant of
the substrate medium $\eps (\omo) < 0$ and the substrate does not
support bulk photon modes. This is the situation that is most
easily implementable in the vicinity of a metallic surface and
which, in fact, has recently received a considerable attention in
the context of both organic excitons \cite{bellesa2004} and dipole
excitations of nanoparticles (Ref.~\onlinecite{bozhe2004} and
references therein). The substrate in general affects both the
dispersion of the excitons and their life-time. We will show that
the presence of the substrate may result in new interference
patterns and leads to a plethora of behaviors depending on the
polarization and frequency of the excitation as well as on the
distance from the interface. For the decay width, this is
illustrated in Fig.~\ref{im_10_0} that exemplifies substantial
differences with the vacuum result (\ref{radvac}) and is discussed
in more detail later. This Figure demonstrates not only a
modification of the exciton decay into vacuum photons but also the
decay into substrate surface plasmon (SP) modes (that occur for
$\eps (\omo) < -1$), significance of the latter channel strongly
increasing upon approach to the interface.

\begin{figure*}
\includegraphics[scale=0.75]{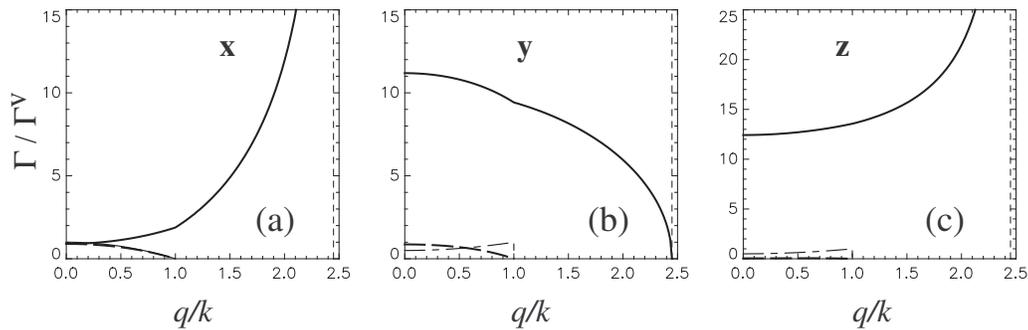}
\caption{\label{im50_1_2}As in Fig.~\ref{im_10_0} but for $d=50\,
a$ and $\eps=-1.2$. Note that despite the separation distance here
is 5 times larger than in panels (d-f) of Fig.~\ref{im_10_0}, the
decay into substrate surface plasmons is much stronger.}
\end{figure*}

Importance of the coupling of an individual dipole excitation to
substrate SP  modes and of associated resonance decay and
scattering phenomena have already been stressed both for molecules
\cite{morphil1974,phil1975a,weber1979} and nanoparticles.
\cite{bozhe2004} This coupling strongly increases as the exciton
transition frequency approaches the ``resonance'' region of the
substrate in which $\eps (\omo)$ is close to $-1$. The increase
is, of course, what should be expected from the theory of
electrostatic image forces \cite{jackson} which features the
combination factor
\begin{equation}\label{Q0}
\Qo=(1-\eps)/(1+\eps)
\end{equation}
for the magnitude of image charges. One should be aware, however,
that this is also the region where both retardation
\cite{morphil1974,phil1975a} and dissipation \cite{cps1978}
effects are particularly important. The corresponding enhancement
of the decay of 1D excitons into SPs is seen in the illustration
of Fig.~\ref{im50_1_2} ($\eps=-1.2$) where, in comparison with
Fig.~\ref{im_10_0} ($\eps=-10$), it clearly is a dominant decay
channel; consequently the fluorescence efficiency is greatly
reduced.\cite{phil1975a} One also appreciates the fact that the
resonant enhancement, as shown, occurs over the already
``super-radiant'' vacuum decay rate.

\begin{figure}
\includegraphics[scale=0.75]{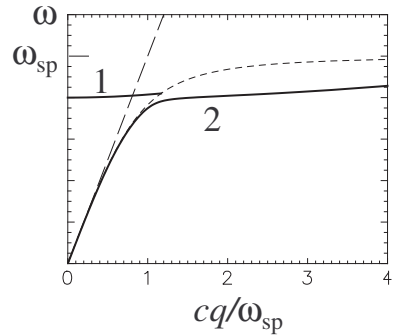}
\caption{\label{schem}Schematically (not to scale!), an example of
the possible idealized dispersion of $\vx$-polarized 1D
exciton-polaritons in the form of their frequency
$\omo=E(q)/\hbar$ as a function of the wave vector $q$. Here the
exciton-polariton spectrum (solid lines) is split into two
branches: the states of branch 1 decay via emission of photons
and/or SPs, the states of branch 2 are non-emissive. This spectrum
is a result of the interaction of the bare excitons with vacuum
photons, whose spectrum is reflected in the limiting long-dash
line, and with substrate surface plasmons, whose spectrum is
reflected in the limiting short-dash line. $\omo_{\mathrm{sp}}$ is
the asymptotic SP frequency.}
\end{figure}

Both Figs.~\ref{im_10_0} and \ref{im50_1_2} have been calculated
with negligible exciton scattering and substrate losses (see
Sec.~\ref{secdiss} for discussion of dissipation effects), hence
they are reflective of the full conservation laws for our
exciton-photon-SP system. So excitons with wave vectors $|q| > k$
cannot decay into vacuum photons, while emission of SPs can occur
only for $|q| < \kvp (k)$ where
\begin{equation}\label{SPP}
\kvp= k \,[\eps/(\eps+1)]^{1/2}
\end{equation}
is the well-known (e.g., Ref.~\onlinecite{Novbook}) wave vector of
the SP at an appropriate frequency. In such an idealized system,
exciton-polaritons with larger wave vector magnitudes would be
non-emissive. An example of the corresponding qualitative picture
of the dispersion spectrum of exciton-polaritons in the chain is
shown schematically in Fig.~\ref{schem}. The non-emissive branch 2
there can be thought of as of excitations representing a coherent
mix of the exciton, photons and SPs of the same momentum
projection along the chain, the relative weight contributions
depending on this momentum. (The branch splitting exhibited in
Fig.~\ref{schem} would not take place for $\vy$-polarized excitons
as consistent with the SP polarization.) We cannot exclude that
surface plasmon guiding by chains of nanoparticles found in recent
numerical simulations \cite{bozhe2006} is related to the formation
of the discussed bound exciton-SP states.

As the SPs are surface states characterized by 2D wave vectors,
the inverse square-root singular behavior in the decay rates of 1D
excitons upon $q \rightarrow \kvp$ seen in Figs.~\ref{im_10_0} and
\ref{im50_1_2} has the ``dimensionality'' origin similar to the
one taking place  for 2D (quantum well) excitons decaying into 3D
vacuum photons \cite{VMAbook,AD,bassani1990,jorda1993} and 1D
(quantum wire) excitons decaying into 2D cavity photons.
\cite{chen2001}

In what follows we will elaborate on the interaction of 1D
excitons with the neighboring reflecting substrate using the
simplest model of a chain of molecules with a single molecular
dipole transition (Frenkel-like excitons). We note that some
resulting features would be generically valid for Wannier-Mott
excitons as well. We however do not pursue here an explicit
analysis of the Wannier-Mott excitons that would require specific
modifications for their bare dispersion as well as the influence
of the substrate on the exciton binding and oscillator strength.
While we chose, for certainty, to use an example of the metallic
substrate and the corresponding surface plasmon excitations, it
should be clear that the same effects would occur for a dielectric
substrate whenever it can feature the region of negative $\eps
(\omo)$ and the surface polariton excitations.

\section{\label{vacuum}Frenkel exciton-polaritons in a molecular chain. Vacuum results.}

Renormalization of a single-molecule transition of energy $\Ef$
and the formation of an electric-dipole exciton band in a
molecular aggregate can be derived in various frameworks (see,
e.g., Refs.~\onlinecite{VMAbook,craig}) with identical results.
Here we will use a simple and physically transparent description
having a clear underlying semi-classical analogy. In the
Heitler-London approximation (energy $\Ef$ is much larger than all
other energies involved), the exciton Hamiltonian for interacting
identical molecules can be written as
\begin{eqnarray*}
H & = & (\Ef+\Vo)\sum_{n}\Bnd\Bn + \sum_{n > m} V_{n-m} (\Bnd\Bm +
\mathrm{h.c.})  \\
& = & \sum_{q} \Eq \Bqd\Bq,
\end{eqnarray*}
where $\Bnd$ creates an excitation on the $n$th molecule whereas
$\Bqd=N^{-1/2}\times$ $\sum_{n}\mathrm{e}^{iqna}\,\Bnd$ creates an
excitation of the wave vector $q$ in a chain of $N$ molecules with
the intermolecular spacing $a$. Correspondingly, the exciton
energy
\begin{equation}\label{en1}
\Eq=\Ef + \Vo + 2 \sum_{n > 0} V_{n} \cos(qna),
\end{equation}
where $\Vo$ represents a possible renormalization for a single
molecule\footnote{Single molecule renormalization is, to a large
extent, not important for our interests in this paper and we refer
the reader to numerous studies dedicated to it. We will completely
disregard the vacuum real energy shift of a single molecule which
can be thought of as absorbed in $\Ef$.} while $V_{n}$ the
\textit{effective} intermolecular interaction as mediated by the
electromagnetic field. The interaction is handily expressed via
the semi-classical expression
\begin{equation}\label{v1}
V_{n}= -\vp\cdot\vE(na\vxu),
\end{equation}
with $\vp$ being the molecular transition dipole moment and
$\vE(na\vxu)$ the electric field produced by the dipole $\vp=p\,
\vpu$ at the distance $na$ along the chain axis chosen to be along
$x$ (we use caps to denote unit vectors).

If the electric field $\vE$ is the total retarded field, however,
it is in fact also a function of the oscillating dipole frequency
$\omo=c k$ and has both real and imaginary parts. Equation
(\ref{en1}) then has to be rewritten as a more general equation
\begin{equation}\label{en2}
E(q)=\Ep - i\Gamma/2=\Ef + \Sigma(k,q),
\end{equation}
involving the self-energy correction $\Sigma(k,q)$, the real-space
transform of which
\begin{equation}\label{v2}
\Sigma(k,x)= -\vp\cdot\vE(k, x \vxu)
\end{equation}
serves to replace (\ref{v1}). Equation (\ref{en2})
self-consistently (via $k=E/\hbar c$) determines both the
dispersion $\Ep (q)=\Ef + \Rep\{\Sigma(k,q)\}$ and the decay width
$\Gamma (q)=-2\,\Imp\{\Sigma(k,q)\}$ of the exciton-polariton
states as a function of their wave vector $q$. This simple
approach, alternatively formulated in terms of Green's functions,
is both physically appealing and powerful as it involves only
classically calculable electric fields; various aspects of it have
been used for different geometries (e.g.,
Refs.~\onlinecite{cps1978,Novbook,hartman2001} and references
therein). In this paper we will not pursue solving specific
self-consistent problems that may depend on a multitude of
parameters and, instead, be discussing the self-energy for a given
\textit{real} value of parameter $k$ (that is, the oscillating
frequency) for different values of the excitation wave vector $q$.
For our numerical illustrations in this paper we chose a
representative value of $k$ corresponding to the wavelength
$\lambda=2\pi/k=400\, a$, a reasonable magnitude for the optical
region of the spectrum given typical spacing $a \sim 10$ \AA.
Understandably, typical values for nanoparticle systems would be
different. \cite{marksar} If neglecting the retardation effects
(purely electrostatic fields), the value of $k$ would have to be
set equal to zero.

Let us briefly review the application to a molecular chain in
vacuum (see also Refs.~\onlinecite{phil1975b} and
\onlinecite{marksar}). Consider the standard \cite{jackson}
retarded electric field at the point $\vrr=r\,\vru$ from the
oscillating point dipole in vacuum:
\begin{eqnarray}
\vEvac(k, \vrr) & = & \frac{\mathrm{e}^{i\ko r}}{r}\left\{ \ko^2
\left[\,\vp -\vru(\vru\cdot\vp) \right] \right. \nonumber \\
& + & \left. \left(\frac{1}{r^2}-\frac{i\ko}{r}
\right)\left[\,3\vru(\vru\cdot\vp)-\vp\right] \right\}
\label{vac1}
\end{eqnarray}
and the corresponding
$$\Svac (k,\vrr)=-\vp\cdot\vEvac(k, \vrr).$$
This expression turns out to be directly applicable even for the
decay of a single molecule (see also Ref.~\onlinecite{Novbook}):
indeed, calculating $-2\,\Imp\{\Svac(k, \vrr\rightarrow 0)\}$
immediately leads to the well-known decay width (\ref{radmol}). To
derive the decay rate for a 1D exciton state with wave vector $q$,
one augments this decay by the sum of contributions from other
molecules:
$$
\Gvac (k,q)= \Go (k) - 4 \sum_{n > 0} \cos(qna)\, \Imp\{\Svac(k,
na \vxu)\}
$$
resulting, after evaluation of the sum, in already quoted
Eq.~(\ref{radvac}). Exemplifying a general feature of self-energy
corrections, a direct inspection easily verifies that, for a fixed
$k$, Eqs.~(\ref{radvac}) and (\ref{radmol}) satisfy, as expected,
\begin{equation}\label{sumrule}
\sum_{q}\Gvac(k,q)=N \, \Go(k).
\end{equation}

With the real part of the field (\ref{vac1}), one immediately
obtains the effective resonant interaction matrix element:
\begin{eqnarray}
\frac{1}{p^2} \,\Rep\{\Svac(k,x)\} & = &
\left(1-3\cst\right)\left[\frac{\cos(\ko
x)}{x^3}+\frac{\ko\sin(\ko x)}{x^2} \right] \nonumber \\ & - &
\left(1-\cst\right)\,\frac{\ko^2\cos(\ko x)}{x}, \label{va2}
\end{eqnarray}
exactly the same result that would be derived in the picture of
the virtual photon exchange. \cite{craig} Calculating the
corresponding sums with $n > 0$ (and disregarding the irrelevant
single-molecule renormalization\footnotemark[\value{footnote}])
for many molecules in the long wavelength, $q a \ll 1$, expansion,
one arrives at
\begin{eqnarray}
\frac{a}{p^2} \,\Rep\{\Svac(k,q)\} & \simeq &
\frac{\left(1-3\cst\right)}{2} \left[
\frac{4\zeta(3)}{a^2}+\ko^2-3q^2  \right. \nonumber \\
& - & \left. \left(\ko^2-q^2\right)b \right] +
\left(1-\cst\right)\,\ko^2 b, \hspace{10mm} \label{vb2}
\end{eqnarray}
where $\zeta(3) \simeq 1.202$ and $b=\ln
\left(|\ko^2-q^2|a^2\right)$. (It is useful to note that
approximation (\ref{vb2}) actually works very well over a sizable
portion of the exciton Brillouin zone.) The known logarithmic
divergence in Eq.~(\ref{vb2}) upon $q\rightarrow k$ signifies the
splitting of the exciton-polariton spectrum into two branches
\cite{AD,phil1975b,citrin1992} as caused by the radiative zone
component of the field at $\ct \neq 1$ in Eq.~(\ref{va2}). Of
course, corresponding to this divergence there is a non-vanishing
decay rate at $q\rightarrow k$ in Eq.~\ref{radvac}. The vanishing
of the latter takes place only at $\ct = 1$, and in this case the
exciton dispersion exhibits a single-branch behavior with a cusp
at $q=k$.

The electrostatic part of Eq.~(\ref{vb2}) features a non-analytic
behavior $\propto q^2\ln(qa)^2$ at $|q| \gg k$ due to the
long-range nature of the dipole-dipole interaction making the
exciton dispersion ``steeper'', the behavior that recently
attracted attention in the context of exchange-interaction effects
for excitons in carbon nanotubes. \cite{perebeinos} The overall
width of the bare exciton zone as seen in Eq.~(\ref{vb2}) is
scaled with the energetic parameter
\begin{equation}\label{enscale}
J=p^2/a^3
\end{equation}
establishing the unit for the nearest-neighbor electrostatic
dipole-dipole interaction. To appreciate the scale of energies
involved: with $p=1$ Debye and $a=10$ \AA, for instance, $J \simeq
0.014$ eV. With reasonable variations of values of $p$ and $a$,
$J$ could reach magnitudes $\sim 0.1$ eV.

\section{\label{Interface}Molecular chain near the interface.}

In the vicinity of the interface with a metallic/dielectric body,
the total electric field can be conveniently represented as a sum
of the primary, vacuum, field, discussed in section \ref{vacuum},
and the secondary field due to the induced response of that body:
$\vE=\vEvac + \vEs$. Correspondingly, the self-energy of the
exciton-polaritons is also represented as $\Vn=\Vnvac + \Vns$. In
what follows we discuss the induced contribution $\Vns (k,q)$
coming from a half-space characterized by the dielectric constant
$\eps=\eps(\omo)$ taken at the frequency $\omo=c k$ in the
geometry of the molecular chain (along $\vx$) being parallel to
the separating interface ($xy$-plane) at the distance $z_{0}=d/2$
from it ($d$ is the distance between a dipole and its image).

The problem of an electric dipole near a metallic/dielectric
half-space is a classical problem first treated by Sommerfeld
\cite{Sombook,Novbook} and whose solution is available in
different forms. Here we find it convenient to adopt the
expressions for the electric field as derived in
Ref.~\onlinecite{Kingbook}. In the context of our application for
fields along the chain, it matters how the dipoles are oriented
with respect to both the chain and the interface. For a chain of
electric dipoles of an arbitrary polarization $\vpu$, one easily
finds that
$$\Wxs =\puxt\Vxsx+\puyt\Vxsy+\puzt\Vxsz,$$
where axes-related components can be rewritten from results in
Ref.~\onlinecite{Kingbook} as follows:
\begin{equation}\label{xpot}
\frac{1}{p^2}\,\Vxsx=\int_{0}^{\infty} \kv\, d\kv\,
\textrm{e}^{-\ga d} \left( \frac{\ga Q}{2} J_{-}(\kv x) +
\frac{\ko^2 P}{2\ga} J_{+}(\kv x) \right)
\end{equation}
for $\vx$-polarized dipoles,
\begin{equation}\label{ypot}
\frac{1}{p^2}\,\Vxsy=\int_{0}^{\infty} \kv\, d\kv\,
\textrm{e}^{-\ga d} \left( \frac{\ga Q}{2} J_{+}(\kv x) +
\frac{\ko^2 P}{2\ga} J_{-}(\kv x) \right)
\end{equation}
for $\vy$-polarized dipoles, and
\begin{equation}\label{zpot}
\frac{1}{p^2}\,\Vxsz=\int_{0}^{\infty} \kv\, d\kv\,
\textrm{e}^{-\ga d} \ \frac{\kv^2 Q}{\ga} J_{0}(\kv x)
\end{equation}
for $\vz$-polarized dipoles. In Eqs.(\ref{xpot}-\ref{zpot}),
$$J_{\pm}(x)=J_{0}(x)\pm J_{2}(x)$$
are composed of Bessel functions of the first order while
parameters
\begin{equation}\label{QP}
Q=\frac{\gae  - \eps \ga}{\gae  + \eps \ga}, \ \ \ \
P=\frac{\gae-\ga}{\gae+\ga}
\end{equation}
and
\begin{equation}\label{gammas}
\ga=(\kv^2-\ko^2)^{1/2}, \ \ \gae=(\kv^2-\eps\ko^2)^{1/2}
\end{equation}
(for negative $u < 0$, $u^{1/2}=-i(-u)^{1/2}$ should be used in
Eq.~(\ref{gammas}).) One can straightforwardly verify that the
no-retardation limit ($\ko=0$) of the above expressions leads to
usual electrostatic fields of image dipoles.

\begin{figure*}
\includegraphics[scale=0.75]{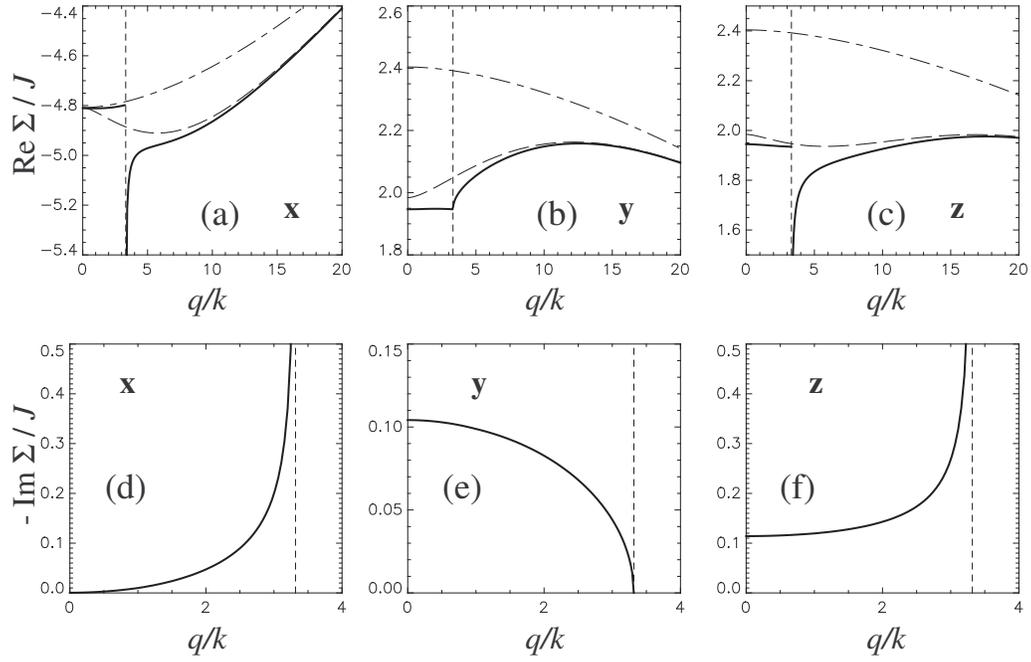}
\caption{\label{full10_1_1}Real (upper panels) and imaginary
(lower panels) of the self-energy $\Sigma (k,q)$ of
exciton-polaritons as functions of the reduced wave vector $q/k$
for a fixed value of $k=2\pi/400\, a$, $\eps=-1.1$ and $d=10 a$.
Self-energy is shown with respect to $J$ from Eq.~(\ref{enscale}).
The vertical short-dash lines show the position of the surface
plasmon wave vector, $\kvp/k$. In the upper panels, the long-dash
lines show the electrostatic (no-retardation) results, dash-dotted
lines the electrostatic results for the exciton dispersion in
vacuum.}
\end{figure*}

Representations (\ref{xpot}-\ref{zpot}) are quite meaningful
physically. One immediately observes that the pole in the
parameter $Q$ in (\ref{QP}) occurs at $\kv$ equal to $\kvp$ in
Eq.~(\ref{SPP}), the wave vector of the surface plasmon (surface
polariton) whenever real $\eps < -1$. Together with $p$-polarized
photons, SPs make the $Q$-containing contributions in
Eqs.~(\ref{xpot}-\ref{zpot}). The parameter $P$-containing terms,
on the other hand, correspond to contributions to the electric
fields from $s$-polarized photons. Expressions
(\ref{xpot}-\ref{zpot}) taken at the source point $x=0$ would
describe the effect of the half-space on the electronic transition
in a single molecule as studied in
Refs.~\onlinecite{morphil1974,phil1975a}.

We now turn to the $q$-dependent self-energy of an exciton in a
long chain of molecules. Restricting to the chain-to-interface
distances larger than the intermolecular spacing: $d \gg a$, one
can safely use a continuum description of the half-space response:
\begin{equation}\label{cont}
\Dqs=2\int_{0}^{\infty}\frac{dx}{a}\, \Vxs \cos(qx).
\end{equation}
As with the vacuum case, all results are clearly even functions of
$q$; to simplify expressions, we will therefore continue assuming
$q > 0$. Transformation (\ref{cont}) for individual dipole
contributions (\ref{xpot}-\ref{zpot}) is facilitated by the
integrals: \cite{Prud}
\begin{equation}\label{int1}
\int_{0}^{\infty} dx\, \cos(qx)J_{0}(\kv x) =
\frac{1}{(\kv^2-q^2)^{1/2}}\, \Theta(\kv-q),
\end{equation}
\begin{equation}\label{int2}
\int_{0}^{\infty} dx\, \cos(qx)J_{2}(\kv x) =
\frac{1-2q^2/\kv^2}{(\kv^2-q^2)^{1/2}} \, \Theta(\kv-q)
\end{equation}
so that Eq.~(\ref{zpot}), e.g., is transformed into
\begin{equation}\label{zpotq}
\frac{a}{p^2}\,\Vqsz = 2 \int_{q}^{\infty}  \frac{\kv^3 Q
\textrm{e}^{-\ga d}}{\ga (\kv^2-q^2)^{1/2}}\, d\kv
\end{equation}
and similarly for Eqs.~(\ref{xpot}) and (\ref{ypot}).
Step-functions in Eqs.~(\ref{int1}) and (\ref{int2}), as is also
reflected in Eq.~(\ref{zpotq}), have a clear physical significance
of the energy-and-momentum conservation limitation.

It is also useful and meaningful to note the no-retardation limit
($\ko=0$) of the above expressions, when $Q$ in Eq.~(\ref{QP})
becomes equal to the electrostatic combination (\ref{Q0}), and the
exciton dispersion would be affected by the image dipoles as
\begin{equation}\label{elx}
\frac{a}{\Qo p^2} \,\Dqsx=2  q^2 \int_{q}^{\infty}
\frac{\textrm{e}^{-\kv d}}{(\kv^2-q^2)^{1/2}}\, d\kv = 2 q^2 \Ko
(qd),
\end{equation}
\begin{equation}\label{ely}
\frac{a}{\Qo p^2} \,\Dqsy=2\int_{q}^{\infty} (\kv^2-q^2)^{1/2}\,
\textrm{e}^{-\kv d}\, d\kv = \frac{2q}{d}\Kone (qd),
\end{equation}
and
\begin{equation}\label{elz}
\Dqsz=\Dqsx+\Dqsy,
\end{equation}
where $\Ko (x)$ and $\Kone (x)$ are the modified Bessel functions.

The effects of these real image corrections are displayed in
Fig.~\ref{full10_1_1} showing an example of the total
$q$-dependent exciton self-energy $\Dkq$ including both vacuum and
secondary field contributions. The figure is a result of an
illustrative calculation for an idealized (no dissipation) system
in the resonance region ($\eps=-1.1$ so that $\kvp \simeq 3.3 k$)
at a relatively close ($d=10\, a$) distance from the interface --
to demonstrate a possible magnitude of the effects. The figure
shows both the retardation effects as well as the fact that upon
the increase of the wave vector $q$, the exciton dispersion
approaches the electrostatic behavior. It is clear that, in
principle, the electrostatic image forces may appreciably modify
the exciton dispersion.

\begin{figure*}
\includegraphics[scale=0.75]{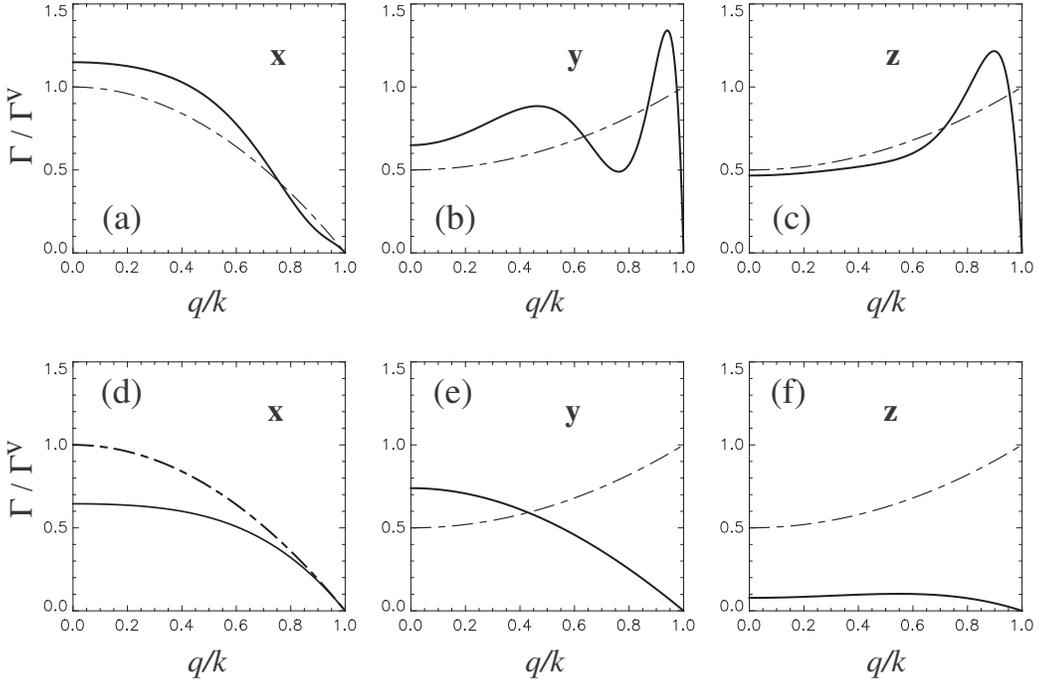}
\caption{\label{im_0_8}Decay width $\Gamma (k,q)$ of the
exciton-polariton as a function of the reduced wave vector $q/k$
for a fixed value of $k$ and $\eps=-0.8$. Two rows correspond to
different dipole-to-image-dipole distances: upper panels (a--c)
are for $d=600 a$, lower panels (d--f) for $d=10 a$. There is no
surface plasmon channel here, and the total decay width is due to
the vacuum photons only.}
\end{figure*}

In an idealized system, the secondary field contributions to the
imaginary part of the self-energy (exciton decay) can come only
from two sources. First, it is the regions of variable $\kv$ in
the integrals where the parameter $\ga$ (\ref{gammas}) is
imaginary. As Eq.~(\ref{zpotq}) shows, that happens only for $q <
k$ -- this is the source of the modification by the substrate of
the exciton decay into vacuum photons that we explicitly
illustrated in Fig.~\ref{im_10_0}. Second, and dominating in
Fig.~\ref{full10_1_1}, source is the surface plasmon pole
$\kv=\kvp$ in the parameter $Q$ in the integrals. As
Eq.~(\ref{zpotq}) shows again, this pole contributes only for $q <
\kvp$ signifying the decay of the exciton into a SP. The pole
contributions to $-\Imp\{\Dkq\}$ are immediately calculable and
can be represented by a combination of a common factor
\begin{equation}\label{p1}
\frac{4\pi }{|\eps|^{1/2}(1-\eps)}\,\frac{p^2 \kvp^4}{a
k^2}\,\exp\left[-d\,(\kvp^2-k^2)^{1/2}\right]
\end{equation}
and polarization- and $q$-dependent co-factors:
\begin{equation}\label{p2}
q^2/\kvp\, (\kvp^2-q^2)^{1/2}
\end{equation}
for $\vx$ polarization,
\begin{equation}\label{p3}
(\kvp^2-q^2)^{1/2}/\kvp
\end{equation}
for $\vy$ and
\begin{equation}\label{p4}
|\eps|\,\kvp/(\kvp^2-q^2)^{1/2}
\end{equation}
for $\vz$.

Accompanying decay's inverse square-root singularity in
Eqs.~(\ref{p2}) and (\ref{p4}), also seen in
Fig.~\ref{full10_1_1}(d) and (f), are the diverging
discontinuities in the real parts (panels (a) and (c) of
Fig.~\ref{full10_1_1}) of the type familiar from the studies of
the decay of 2D excitons into 3D vacuum photons.
\cite{VMAbook,AD,bassani1990,jorda1993} Due to such a
discontinuity, the resulting self-consistent dispersion of, e.g.,
$\vx$-polarized excitations would split into two branches as shown
in Fig.~\ref{schem}. On the contrary, the decay rate of
$\vy$-polarized excitons vanishes at $q \rightarrow \kvp$ in
Eq.~(\ref{p3}). This is a consequence of the polarization of SPs
whose electric field can have only longitudinal and perpendicular
to the interface components (like $p$-polarized photons). As $q
\rightarrow \kvp$, the surface plasmons would be emitted along the
chain direction, hence their field would have no $\vy$-components
to interact with $\vy$-excitons. (There is a similarity here with
the decay of $\vx$-polarized excitons in vacuum, whose decay rate
vanishes at $q \rightarrow k$, Eq.~(\ref{radvac}).)
Correspondingly, the real part of the self-energy in
Fig.~\ref{full10_1_1}(b) does not exhibit a discontinuity at
$q=\kvp$ (the discontinuity is in the derivative) with the
resulting exciton spectrum consisting of one branch only. Quite
clear is also $\propto q^2$ behavior for $\vx$-excitons in
Eq.~(\ref{p2}), it has the same origin as in the electrostatic
effect (\ref{elx}) -- at $q \rightarrow 0$, the exciton
polarization of the chain becomes uniform and there would be no
polarization charges (a vanishing spatial derivative) to induce
images in the substrate.

The SP excitations in the substrate exist in the region of
frequencies where $\eps (\omo) < -1$. Qualitatively different,
therefore, for a reflecting substrate is another region of
frequencies in which $-1 < \eps (\omo) < 0$. In this case
exciton-polaritons in the molecular chain can decay only into
vacuum photons. The (idealized) substrate then serves as to modify
the dispersion of excitons and their radiative decay. An example
of the radiative decay modification is shown in Fig.~\ref{im_0_8}
calculated for $\eps = -0.8$ and to be compared with
Fig.~\ref{im_10_0}. Both figures feature the same distances from
the interface but different magnitudes \textit{and} signs of the
image charges, Eq.~(\ref{Q0}). These two facts explain
similarities and differences in the radiative decay patterns for
the two examples. Among the important common features, one should
notice ``oscillating'' $q$-dependences of the decay rate at larger
distances and the vanishing of the radiative decay at $q
\rightarrow k$. As the chain is moved further away from the
interface, even more undulations would be observed in the
$q$-dependence of the decay rate as a result of the interference
with the image dipoles.

\section{\label{secdiss}Effects of dissipation in the substrate}

\begin{figure*}
\includegraphics[scale=0.75]{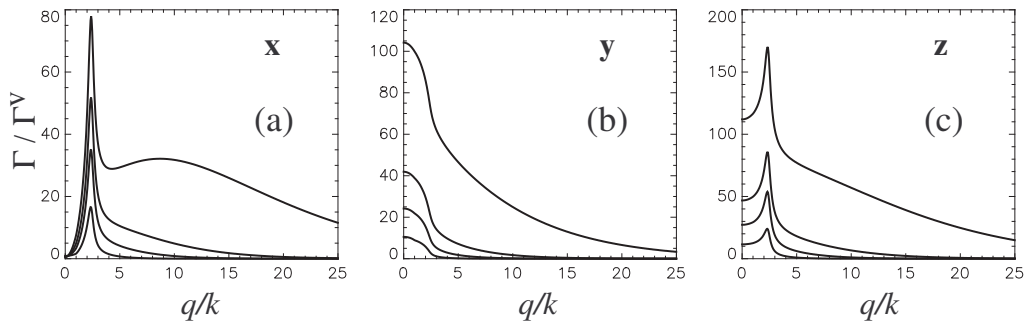}
\caption{\label{fam_1_2_0_05}Decay width $\Gamma (k,q)$ of
exciton-polaritons as a function of the reduced wave vector $q/k$
for a fixed value of $k=2\pi/400\,a$ in units of vacuum
$\Gamma^{\mathrm{v}}=2\pi p^2 k^2/a$ and for different (indicated
in the panels) exciton polarizations. Here substrate's dielectric
constant is \textit{complex}: $\eps=-1.2 + 0.05 i$. Each of the
panels contain 4 curves corresponding to different distances
between the molecular chain and the substrate (top to bottom):
$d=10\,a$, $20\,a$, $30\,a$ and $50\,a$.}
\end{figure*}
\begin{figure*}
\includegraphics[scale=0.75]{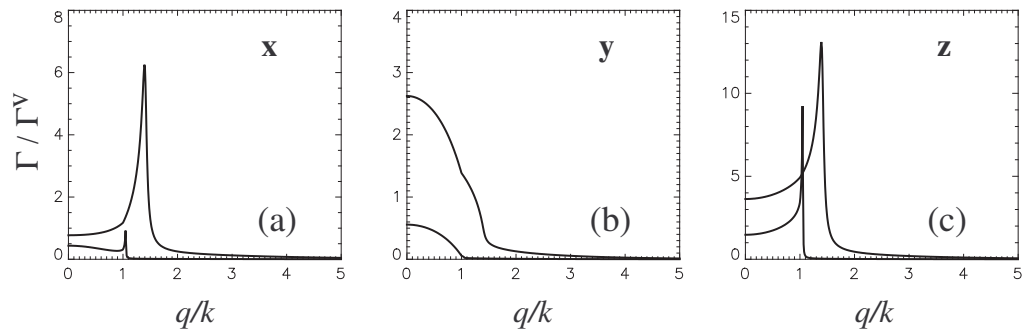}
\caption{\label{fam_50a}As in Fig.~\ref{fam_1_2_0_05} but for
$d=50\, a$ and two different dielectric functions (top to bottom):
$\eps=-2.0 + 0.1 i$ and $\eps=-10.0 + 0.5 i$.}
\end{figure*}

Real metallic (dielectric) substrates are characterized by finite
dissipation (losses) that are taken into account
phenomenologically via the imaginary part of the dielectric
function:
$$
\eps (\omo)=\epsp (\omo) + i\epspp (\omo).
$$
In the context of the effects of $\epspp$ on SPs, for a given real
frequency $\omo$, wave vectors of the SPs consequently acquire
imaginary parts as well:\cite{Novbook}
$$
\kvp=\kvpp + i\kvppp,
$$
where $\kvppp$ describes the damping of the SP modes. Another way
to look at $\kvppp$ is as of the uncertainty (broadening) of the
SP's wave vector. Hence the momentum conservation law in the
exciton decay into a SP would not be obeyed exactly as in
idealized systems discussed so far. One thus should expect a
broadening and extension of the range of finite exciton decay
rates into the region of the exciton wave vectors $q > \kvpp$.
Moreover, even with \textit{relatively} small losses, $\epspp \ll
|\epsp|$, the relative damping of the SP can become substantial in
the resonance region of $\epsp$ close to $-1$. Indeed, in this
case\cite{Novbook}
\begin{equation}\label{compl3}
\kvppp/\kvpp \simeq \frac{\epspp/\epsp}{2(\epsp+1)},
\end{equation}
and the denominator in the r.~h.~s. of Eq.~(\ref{compl3}) could
``compensate'' for the smallness of $\epspp$ and completely
destroy the notion of a coherent SP.

Figure \ref{fam_1_2_0_05} shows results of the calculation of the
exciton decay for the substrate with the dielectric constant
$\eps=-1.2 + 0.05 i$ and variable distances to the interface. As
compared to the idealized system, one immediately notices very
broad, especially at smaller distances, $q \gg \kvpp$ regions of
the exciton decay. For this particular value of the complex
dielectric function, however, the ``uncertainty'' ratio
(\ref{compl3}) is only about 0.1. Thus the extremely broadened
$q$-region of the exciton decay is not due to the broadening of
the SPs themselves, the latter can explain only the formation of
the finite magnitude SP peaks clearly seen in
Fig.~\ref{fam_1_2_0_05}. Of course, with complex dielectric
constants, the imaginary part of the exciton self-energy discussed
in Sec.~\ref{Interface} is contributed to, in principle, by the
whole integration range in integrals like Eq.~(\ref{zpotq}),
rather than just by the around-the-pole region. What is reflected
in the broad ``wings'' in Fig.~\ref{fam_1_2_0_05} is the result of
the ordinary ``incoherent'' Joule losses due to the oscillating
electric field of the exciton polarization penetrating into the
substrate. Especially illuminating in this regard is a second
maximum on the top-most curve in Fig.~\ref{fam_1_2_0_05}(a).
Indeed, similarly to the already discussed in Sec.~\ref{Interface}
image-charge effects, the electrostatic component of the electric
field of $\vx$-polarized excitons has a $q$- and $d$-dependence
reflected in the r.~h.~s. of Eq.~(\ref{elx}), the maximum of which
is achieved at $q \sim 1/d$. As the distance from the interface
increases, the role of such short-range energy transfer from the
higher-$q$ excitons to the substrate decreases and one can
``tune'' it off while still having an appreciable decay rate of
the lower-$q$ states into SPs. In addition to the largest-distance
curves in Fig.~\ref{fam_1_2_0_05}, this point is illustrated in
Fig.~\ref{fam_50a} showing clear SP emission effects both close to
and away from the resonance region. The distance dependence of the
SP emission intensity is governed by the factor (\ref{p1}) and one
could likely ``optimize'' the relationship between the coherent
and incoherent energy transfer based on the system regime
parameters.

\section{Discussion.}

The modification of the electric fields in the presence of the
reflecting substrate leads to substantial changes of the
properties of 1D exciton-polaritons in a dipole-coupled chain
above the substrate. The substrate modifies the interaction of the
exciton with vacuum photons and, in addition, can engage a new and
efficient interaction channel -- with surface excitations of the
substrate. While these statements also apply to a
well-studied\cite{morphil1974,phil1975a} case of a single molecule
near the interface, the cooperative interaction of many molecules
in the chain and of their images may lead to qualitatively and
quantitatively interesting effects. In this paper, we discussed
such effects for coherent delocalized 1D exciton-polaritons that
are well characterized by their wave vectors $q$. We emphasize
that it is actually individual $q$-states that would exhibit
qualitatively different coherent properties -- if averaged
uniformly over all $q$, a (near) reduction would occur to the
effects of the substrate on an individual molecule (similarly to
discussed Eq.~(\ref{sumrule}) for vacuum). Immediately noticeable
in our illustrative examples, for instance, is a difference of the
behavior of $\vx$- and $\vy$-polarized 1D exciton-polaritons --
while those cases of the molecular transition dipoles parallel to
the interface would be equivalent for single molecules. Properties
of 1D exciton-polaritons depend specifically on their
polarization, frequency with respect to the substrate dielectric
dispersion and distance from the interface. We reiterate that,
while our illustrations used parameters appropriate for molecular
systems, qualitatively similar effects are expected for other
systems such as chains of nanoparticles.

A common feature for all exciton polarizations and conditions
considered (see Figs.~\ref{im_10_0}, \ref{im50_1_2}, \ref{im_0_8})
is that the rate of the radiative decay into vacuum photons in the
presence of the substrate vanishes at $q=k=\omo/c$, that is, where
the exciton dispersion curve would cross the vacuum photon
dispersion line (see, e.g., Fig.~\ref{schem}) -- this is
distinctly different from the case of the chain in the vacuum
where such vanishing would occur only for excitons polarized
parallel to the chain. The consequence of this effect of the image
dipoles is that no branch splitting occurs at $q=k$. At distances
from the interface comparable to the wavelength $2\pi/k$ of vacuum
photons, the interference with the radiative fields of the image
dipoles results in undulated patterns of the radiative decay
(panels (a-c) of Figs.~\ref{im_10_0}, \ref{im_0_8}) as functions
of $q$; the larger the distance the more undulations would take
place.

It is important to note that, as a result of integration
(\ref{cont}) over many molecules in the chain,  the magnitudes of
these and some other $q$-dependent features fall off with the
distance from the interface slower than their averages
characteristic of single molecules. Consider, e.g., the
electrostatic exciton energy shifts due the interaction of the
molecular dipole(s) with their image(s). For a single molecule,
the corresponding transition energy shift is $\sim p^2/d^3$. The
maximum of $q$-dependent shifts, however, as
Eqs.~(\ref{elx}-\ref{elz}) show, would be $\sim p^2/a d^2$ -- that
is, much larger at distances $d \gg a$ (not in the immediate
proximity of the interface). This may provide an opportunity of
easier experimental identification of such shifts than for
individual molecules.

While the distance dependence (\ref{p1}) of the decay rate of the
excitons into surface plasmons is the same as for single
molecules, $q$-dependent enhancing factors (\ref{p2}) and
(\ref{p4}), seen as the peaks in our illustrations, may facilitate
better experimental verifications. As the rate of the emission of
SPs is $q$-dependent, we speculate that the chain excitons could
perhaps serve as directional sources of SPs -- in expectation of a
more efficient emission along the chain for $x$- and $z$-polarized
excitons and perpendicular to the chain for $y$-polarized
excitons. We recall that these polarization assignments stem from
the fact that the electric field of the SPs lies in the plane made
by the wave vector and the normal to the interface. As the
small-$q$ exciton decay into SPs can be strongly enhanced in
comparison with the decay into vacuum photons, it is likely that
the inverse process of the excitation of the chain by SPs could
also be exploited. Non-emissive exciton-polaritons with larger $q$
may also present an opportunity to be used for SP
guiding.\cite{bozhe2006}

Various scattering and dissipation processes are known to be able
to strongly affect features characteristic of idealized systems.
We particularly discussed the chain exciton quenching due to the
``incoherent'' energy transfer to the substrate -- and this does
not exhaust the list. It suffices to also mention, e.g., the
scattering by phonons in molecular chains or Joule losses in the
chains of metallic nanoparticles. The fast scattering between
different $q$-states would result in the thermalized population of
the exciton-polaritons so that observed decay rates are thermally
averaged (see Refs.~\onlinecite{citrin1992},
\onlinecite{spataru2005}, \onlinecite{perebeinos} for some
specific 1D applications). We therefore presume that the best
conditions to experimentally address finer $q$-dependent effects
we discussed in this paper would be low-temperature spectroscopic
measurements.

\acknowledgments

We gratefully acknowledge support from AFOSR grant FA
9550-05-1-0409 and from the Collaborative U.~T.~Dallas--SPRING
Research and Nanotechnology Transfer Program. VMA also thanks
Russian Foundation of Basic Research and Ministry of Science and
Technology of Russian Federation.

\bibliography{dipoles}

\end{document}